\documentclass[a4paper]{article}
\usepackage{amsmath,amssymb,amsfonts,graphicx}
\long\def\symbolfootnote[#1]#2{\begingroup%
\def\thefootnote{\fnsymbol{footnote}}\footnote[#1]{#2}\endgroup} 

\def\bequ{\begin{equation}}
\def\eequ{\end{equation}}

\def\im{{{\rm i}}}
\def\Tr{{{\rm Tr }}}
\def\R{{{\mathbb R}}}

\def\ie{{\it i.e. }}
\input amssym.def
\input amssym.tex
\def\nn{\nonumber}
\def\ben{\begin{equation}}
\def\een{\end{equation}}

\def\bea{\begin{eqnarray}}
\def\eea{\end{eqnarray}}

\begin{document}

\begin{flushright}
DAMTP-2010-10
\end{flushright}

\begin{center}
{\large {\bf $CP$ Violation Makes Left-Right Symmetric Extensions With Non-Hermitian Mass Matrices Appear Unnatural}}
\\\vspace{15pt}
Steffen Gielen\symbolfootnote[2]{\tt sg452@damtp.cam.ac.uk}
\\{\it DAMTP, University of Cambridge, Wilberforce Road, Cambridge CB3 0WA, UK}
\\\vspace{15pt}
\today
\end{center}

\begin{abstract}
Following a similar recent analysis for $CP$ violation in the electroweak sector of the standard model, we estimate the naturalness of a magnitude of $CP$ violation (measured by the Jarlskog invariant $J$) close to the observed value in extensions of the standard model with left-right symmetry, such as the Pati-Salam model, where quark mass matrices are not Hermitian in general. We construct a simple and natural measure on the space of complex matrices which is both geometrically motivated and uses the observed quark mass hierarchy. We find that, unlike in the case of the standard model where the observed value for $J$ seemed rather typical, one would now expect to observe $|J|< 10^{-7}$, clearly in conflict with the observed value $J\approx 3\times 10^{-5}$. The crucial difference in the calculation lies in the non-Hermiticity of mass matrices that modifies the measure. We conclude that one would need additional assumptions modifying the measure to reproduce the observed value, and that in this sense the standard model is preferred to certain classes of left-right symmetric extensions: It does not need additional assumptions to explain the magnitude of $CP$ violation.
\end{abstract}

\section{Introduction}

In modern theoretical physics one often tries to make statements about ``naturalness" or ``fine-tuning" of the observed values of fundamental parameters, where fine-tuning of a parameter is interpreted as an indication for incompleteness of the theory. Popular examples of fine-tuning problems include the quark mass hierarchy and the cosmological constant problem in particle physics. Since statements about naturalness are fundamentally of a statistical character, to make them mathematically precise one has to assume a well-motivated probability distribution on the parameter space relevant for the theory. A fine-tuning problem then indicates that the probability distribution one has used should be modified by introducing new physical considerations. As an example, as long as observation was consistent with a vanishing cosmological constant, it seemed reasonable to assume that a postulated symmetry would constrain it to vanish. With more recent observations indicating that it must be taken to be very small and positive, there seems to be an issue of fine-tuning\footnote{For an alternative interesting but presumably non-mainstream viewpoint, see \cite{rovellilambda}}. We note that much of the motivation to extend the standard model of particle physics is driven by such considerations, that it is by no means necessary to contemplate a ``Multiverse'' where all possible values of a given parameter are actually realised, and that we need not consider anthropic arguments. Of course we can only ever observe and make measurements in a single Universe, and if a parameter takes a value that appears unlikely maybe this just means that an unlikely possibility is realised in our Universe. One is merely doing statistics. But one should recall Bayes' theorem \cite{bayes}
\ben
P(A|B)=\frac{P(B|A)P(A)}{\sum_{A_i}P(B|A_i)P(A_i)},
\een
where one can take $A$ as a hypothesis and $B$ as an observation. Once we specify {\sc a priori} probabilities for a full set of possible hypotheses $A_i$, and compute the probability of observing $B$ given $A_i$, this allows us to make a statement about the probability of the hypothesis following from the observation. If the priors $P(A_i)$ are modified, one might get a very different answer for $P(A|B)$.

The particular example we will investigate in this paper is the apparently weak magnitude of $CP$ violation in the electroweak sector of the standard model, measured by the Jarlskog invariant $J$. The observation by Kobayashi and Maskawa \cite{kobayashimaskawa} that $CP$ violation is only possible for at least three quark families has led to a Nobel prize, but the issue of possible fine-tuning in the magnitude of $CP$ violation is much less understood.

It is true that, if one is talking about parameters in quantum field theory, they should really not be regarded as constants, but have an evolution with energy scale given by renormalization group equations. However, since fine-tuning means a discrepancy of several orders of magnitude, it may well be that a fine-tuning problem is present at all energy scales. This is true for the quark mass hierarchy \cite{massref}, and also for the case of $CP$ violation: Recent numerical studies \cite{cpviolref} indicate that $J^2$ does not run strongly with energy scale, but that the value at extremely high energies $(\sim 10^{15} {\rm GeV})$ is merely about twice the value at low energies. It is then meaningful to talk about ``naturalness" of the value of such a parameter.

 Since any statements one tries to make depend very directly on the choice of measure, it is helpful if geometric considerations allow for a natural choice of probability distribution. For instance, if the parameter space is a homogeneous space $G/H$ for $G$ a compact Lie group and $H$ a closed subgroup, the natural requirement on the measure determining the probability distribution is invariance under the left action of $G$, which leads to a unique measure (up to normalization). Probabilities for a given function on $G/H$ to take certain values are then well-defined. 

In the case of $CP$ violation in electroweak theory, the parameter space is the space of Cabibbo-Kobayashi-Maskawa (CKM) matrices, a double quotient $H\backslash G/H$, which makes the problem of determining a natural probability distribution more involved than for a homogeneous space $G/H$. This problem was discussed in \cite{paper}, where several possible choices were investigated. It was found that while there is no clearly preferred choice of measure, there always seems to be fine-tuning in the observed value for $J$, unless additional input is used.

In the second part of \cite{paper} (summarised in \cite{prl}), the observed values for the quark masses were taken into account by considering not the space of CKM matrices, but the space of mass matrices, as the fundamental parameter space. This was motivated by the observation that the mass matrices are directly linked to the Yukawa couplings and the Higgs vacuum expectation value, whereas the CKM matrix is only a derived quantity. The observed values for the quark masses were then taken as given, and a probability distribution was constructed that could reproduce these values. The choice made for this distribution was as simple as possible in the following sense: The natural group action on the space of Hermitian mass matrices is the action of the unitary group $U(3)$ by conjugation, $M\rightarrow UMU^{\dagger}$. There is essentially a unique measure invariant under this action. This measure was then modified by introducing the simplest possible function that would allow for a modification of the expectation values for quark masses fitting observation. No further assumptions were needed. It was then found that this simple choice for the measure gave an expectation value for $J$ that was remarkably close to the observed value. Hence the conclusion of \cite{prl} was that once one assumes the quark masses as given, one does not face an additional fine-tuning problem with $J$. This statement, while not new, had been made precise using a geometrically motivated measure on the parameter space.

The calculations done in \cite{paper} heavily relied on the fact that the mass matrices in the standard model can be taken to be Hermitian without loss of generality. Indeed, the starting point was the most natural measure on the space of $3\times 3$ Hermitian matrices. However, in left-right symmetric extensions of the standard model, such as Pati-Salam, such an assumption can no longer be made and the mass matrices have to be regarded as a priori arbitrary complex matrices\footnote{We thank Ben Allanach for pointing this out.}. In this paper, we investigate the consequences for statements about naturalness of $J$ by redoing the analysis of \cite{paper} for general complex matrices. We again use the most symmetric measure on the space of mass matrices, here the space of general complex matrices, and modify it in the simplest possible way to incorporate the observed quark mass hierarchy. While this is a choice that could of course be made very differently, it is a simple choice that uses as few assumptions as possible, and that has worked very well for the case of Hermitian mass matrices. What we find is that the resulting probability distribution is different, and the result is different too: The observed value for $J$ now appears to be unnaturally large, since $CP$ violation should be more heavily suppressed by the quark mass hierarchy. One faces a fine-tuning problem, and needs additional assumptions to modify the measure appropriately.

We should point out that the only real input we use from the physical theory (standard model or a left-right symmetric extension of it) is, apart from the very definition of $CP$-violating parameters, how the theory restricts the type of mass matrices that appear. In particular, for any extension of the standard model (such as left-right symmetric models where parity is the left-right symmetry\footnote{Thanks to the referee for clarification on this point.}) that also has Hermitian mass matrices we would not see any modification in the results. 

The structure of the paper is as follows: In section 2, we review $CP$ violation in the electroweak sector and detail why mass matrices may be assumed to be Hermitian in the standard model, but not if one has an extended left-right symmetry. In section 3, we construct a measure on the space of $3\times 3$ complex mass matrices, taking into account both the invariance of a natural measure under left or right multiplication of a complex matrix by a unitary matrix as well as a decaying function that is necessary for the calculation, includes the observed values for the quark masses, and partially breaks this invariance. In section 4, we compute the expectation value for the square of the Jarlskog invariant $J$ in the given probability distribution, showing that it is much smaller that the observed value. This main part is very similar to the calculations in \cite{paper} and the reader may benefit from comparing with this paper. We summarise in section 5.

\section{$CP$ Violation in the Standard Model and Beyond}

We summarise how $CP$ violation arises, first in the standard model and then in the more general case of left-right symmetric extensions, essentially following \cite{jarlskogbook}. In the standard model, the quark fields appear as left-handed $SU(2)$ doublets and right-handed $SU(2)$ singlets:
\ben
\left(\begin{matrix}q_{jL}\\ q'_{jL}\end{matrix}\right),\quad q_{jR}\,,\quad q'_{jR}\,,\qquad j=1,2,\ldots,N\,.
\een
Here $N$ is the number of quark families, which is arbitrary in the standard model, and normally taken to be three. The fields are written in a flavour basis which can be considered unphysical, since flavour eigenstates do not correspond to mass eigenstates.

The coupling of the Higgs doublet $H$ to quarks, through spontaneous symmetry breaking by the Higgs potential, gives masses to the quark fields:
\ben
\mathcal{L}_{{\rm Higgs}}\stackrel{{\rm SSB}}{\longrightarrow}-\sum_{j,k=1}^N \left(m_{jk} \overline{q_{jL}} q_{kR} + m_{jk}' \overline{q'_{jL}} q'_{kR}\right) +{\rm h.c.}
\label{ssb}
\een
The mass matrices $m$ and $m'$ are determined by the original Yukawa couplings and the Higgs vacuum expectation value. Thus, it seems appropriate to regard either the set of Yukawa couplings together with the Higgs vacuum expectation value or the collection of elements of $m$ and $m'$ as fundamental parameters of the theory. This viewpoint was supported by the results of \cite{paper}. 

This part of the Lagrangian is formally $CP$ invariant if and only if $m$ and $m'$, which are so far arbitrary complex matrices, are real. Since this condition is not satisfied in Nature, one has formal $CP$ violation. However, as remarked before, the Lagrangian has been written in the unphysical flavour basis. One can, for general $m$ and $m'$, pass to a different basis, namely the basis of mass eigenstates, by diagonalising the mass matrices with unitary matrices:
\ben
m=U_L^{\dagger}\Delta U_R\,,\quad m'=(U'_L)^{\dagger}\Delta'U_R'\,,
\label{massdiag}
\een
thus the basis of mass eigenstates is related to the previously considered basis by
\ben
q_{L {\rm phys}}=U_L q_L\,, \quad q_{R {\rm phys}}=U_R q_R\,,\quad {\rm etc.}
\een

It is always possible to choose the unitary matrices so that $\Delta$ and $\Delta'$ are real, and in this new basis this part of the Lagrangian is invariant under $C$ and $P$ separately, and hence also under $CP$. However, the electroweak Lagrangian also contains charged current terms mixing up- and down-type quarks, coupled to the $W$ boson fields via (we are now using the basis of mass eigenstates)
\ben
X_C:=(W_{\mu}^1-iW_{\mu}^2)J_c^{\mu}+{\rm h.c.}\,,\quad J_c^{\mu}:= \overline{(u,c,t)_L}\gamma^{\mu}V\left(\begin{matrix}d_L\\ s_L \\ b_L \end{matrix}\right)\,,
\label{chargedc}
\een
where $V:=U_L (U_L')^{\dagger}$ is the {\it Cabibbo-Kobayashi-Maskawa} (CKM) matrix. In the basis of mass eigenstates, this term $X_C$ is not invariant under $CP$ unless $V$ is real. Since we consider the mass eigenstates as physical, $CP$ is violated through these charged current terms.

An important observation made in \cite{jarlskogframpton} is that one can redefine the right-handed quark bases by arbitrary unitary transformations,
\ben
U_R\rightarrow OU_R\,,\quad U'_R\rightarrow O'U_R'\,,
\label{redefine}
\een
obtaining a new basis which is to be regarded as equally physical. This is due to the absence of charged current terms involving right-handed quarks, since they are singlets under $SU(2)$. It is therefore no loss of generality to set $U_R=U_L$ and $U'_R=U'_L$ in (\ref{massdiag}), and to assume that $m$ and $m'$ are Hermitian. 

A natural way to extend the standard model is to assume the existence of a second $SU(2)$ symmetry which acts on the right-handed quarks, as in the Pati-Salam model \cite{patisalam}. In such extensions, one adds a term (\ref{chargedc}) for right-handed quarks to the Lagrangian. This has the important consequence that a general transformation (\ref{redefine}) for arbitrary unitary transformations $O$ and $O'$ can no longer be regarded as giving an equivalent quark basis, since it modifies this new charged current term. The mass matrices cannot in general be taken to be Hermitian, but are arbitrary complex matrices. 

There are now also two possibly $CP$ violating terms, and two CKM matrices. We focus on $V=U_L (U_L')^{\dagger}$ which involves the processes that are actually observed and disregard $V_R=U_R(U_R')^{\dagger}$ in the following.

Mathematically, $V$ is an element of $SU(N)$, but since the phases of the quark fields are arbitrary (even for non-Hermitian mass matrices), $V$ is only defined up to left or right multiplication by a diagonal element of $SU(N)$, \ie an element of the maximal torus $U(1)^{N-1}$. The space of CKM matrices is therefore the double quotient $U(1)^{N-1}\backslash SU(N)/U(1)^{N-1}$, characterised by $(N-1)^2$ parameters, out of which $\frac{1}{2}N(N-1)$ may be taken to be real (Euler) angles and the remaining $\frac{1}{2}(N-1)(N-2)$ appear as complex phases. It follows that the matrix $V$ can be taken to be real for $N=2$, and so in the formalism explained here one needs at least three quark families to have a possibility of $CP$ violation. Kobayashi and Maskawa were awarded the 2008 Nobel Prize for using this observation to predict the existence of a third quark family \cite{kobayashimaskawa}. We set $N=3$ in what follows.

The mathematical theory of observable measures of $CP$ violation in the standard model was developed by Jarlskog \cite{jarlskogbook}. She showed \cite{jarlsprl} that all necessary and sufficient conditions for $CP$ violation can be summarised as the following condition on the commutator of the Hermitian matrices $m,m'$:
\ben
\det C := \det \left(-{\rm i}\left[m,m'\right]\right)\neq 0\,.
\een
One finds that explicitly
\ben
\det C=-2J(m_t-m_c)(m_c-m_u)(m_u-m_t)(m_b-m_s)(m_s-m_d)(m_d-m_b)\,,
\een
where $J:=\frak{Im}(V_{11}V_{22}V_{12}^*V_{21}^*)$ is the Jarlskog invariant which is invariant under left or right multiplication of $V$ by a diagonal matrix, \ie an element of $U(1)^2$. The geometrical interpretation of the quantity $J$ is given by the so-called unitarity triangles. These express the requirement on $V$ to be unitary, so that for instance
\ben
(VV^{\dagger})_{12}=V_{11}V^*_{21}+V_{12}V^*_{22}+V_{13}V^*_{23}=0\,.
\een
In the complex plane, the three complex numbers that sum to zero form the sides of a triangle. The absolute value $|J|$ is twice the area of this triangle. Since there are different unitarity triangles corresponding to different elements of $VV^{\dagger}$, all with the same area, there are several ways of expressing $J$ in terms of the elements of $V$. A general formula is given by \cite{jarlsprl}
\ben
J\sum_{\gamma, l}\epsilon_{\alpha\beta\gamma}\epsilon_{jkl}=\frak{Im}(V_{\alpha j}V_{\beta k}V_{\alpha k}^*V_{\beta j}^*)\,.
\een
The quantities describing the CKM matrix which are invariant under rephasing of the quark fields are $J$ and the absolute values $|V_{\alpha j}|$.

In the general case where $m$ and $m'$ are not assumed to be Hermitian, the corresponding quantity would be
\ben
\det {\bf C} := \det \left(-{\rm i}\left[m m^{\dagger},m' m'^{\dagger}\right]\right)=-2J(m_t^2-m_c^2)(m_c^2-m_u^2)(m_u^2-m_t^2)(m_b^2-m_s^2)(m_s^2-m_d^2)(m_d^2-m_b^2)\,,
\een
which of course leads to the same conditions on the mass matrices as before (given that we took all masses to be positive before). In the literature, the use of ${\bf C}$ is perhaps more common than the use of $C$, and one may well argue that this second measure of $CP$ violation should be considered more fundamental as its value does not depend on the arbitrary signs of the mass terms in the Lagrangian.

One common parametrization of the CKM matrix, which we use in the following, is given by
\ben
V = \left( \begin{matrix} \cos y \cos z & \cos y \sin z & e^{-{\rm i}w} \sin y \\ -\cos x \sin z - e^{{\rm i}w}\sin x \sin y \cos z & \cos x \cos z - e^{{\rm i}w} \sin x \sin y \sin z & \sin x \cos y \\ \sin x \sin z - e^{{\rm i}w} \cos x \sin y \cos z & - \sin x \cos z - e^{{\rm i} w} \cos x \sin y \sin z & \cos x \cos y\end{matrix} \right)\,,
\een
where the ranges of the Euler angles $x, y, z$ and the complex phase $w$ are
\ben
0 \le x, y, z \le \frac{\pi}{2}\,, \quad 0 \le w < 2\pi\,.
\een
An arbitrary $SU(3)$ matrix can then be written as
\ben
U = T_L\,V\,T_R\,,
\een
where
\ben
T_L = {\rm diag}(e^{2{\rm i}p},e^{{\rm i}(q-p)},e^{-{\rm i}(p+q)})\,,\quad T_R = {\rm diag}(e^{{\rm i}(r+t)},e^{{\rm i}(r-t)},e^{-2{\rm i}r})\,,
\een
and $p,q,r$, and $t$ are phases which can take all values between $0$ and $2\pi$. This shows that the coordinates $x,y,z$, and $w$ indeed parametrise representatives of the double quotient $U(1)^2 \backslash SU(3) / U(1)^2$.

In this parametrization, the Jarlskog invariant $J$ is given by
\ben
J = \frac{1}{4} \sin 2x \sin 2z \sin y \cos^2 y \sin w\,.
\een

It appears that the observed value for $J$ is very small, since the maximal value would be $\frac{1}{6\sqrt{3}}\approx 0.1$, whereas in Nature \cite{databook}
\ben
J=3.05_{-0.20}^{+0.19}\times 10^{-5}\,.
\label{jobserved}
\een

In a general discussion where the values of the quark masses are not fixed, $J$ is not an appropriate measure of $CP$ violation, since even with nonvanishing $J$ one could have $CP$ conservation if, for example, the up and charm quark masses were coinciding. It was suggested in \cite{jarlskog2} to use an appropriately normalised form of $\det C$, namely
\ben
a_{CP}= 3\sqrt{6}\frac{\det C}{(\Tr\,C^2)^{3/2}}
\een
for three quark families as the unique basis independent measure of $CP$ violation. This is a dimensionless number which takes values between $-1$ and $+1$, and is again observed to be very close to zero. When written out in terms of the CKM matrix parameters and quark masses, it is a rather complicated expression which is therefore not extremely useful in practical computations. In the present analysis, we assume the quark masses as known and regard $J$ as the measure of $CP$ violation. 

\section{Measure on the Space of $3\times 3$ Complex Matrices}

In this section we determine a natural measure on the space of $3\times 3$ complex matrices in order to make statements about likely or natural values for the magnitude of the Jarlskog invariant $J$. Following \cite{paper}, this measure is a product of the geometrically most natural measure with maximal symmetry and a factor that involves the observed values for the quark masses. This second factor is introduced for two reasons; firstly, to make the total volume of the parameter space finite, secondly, to allow for a modification of the distribution on the space of CKM matrices through the quark masses. The analysis is analogous to the case of $3\times 3$ Hermitian mass matrices discussed in \cite{paper}, and we will see shortly where differences arise that eventually lead to different results.

The space of $3\times 3$ complex matrices has a natural left and right action by $U(3)$ corresponding to changes of basis. We first determine a metric invariant under these group actions. We start with Jarlskog's representation of an arbitrary complex matrix as
\ben
M=U_L^{\dagger}DU_R\,,
\een
where $U_L,U_R$ are unitary and $D={\rm diag }(D_1,D_2,D_3)$ is real diagonal. Here, as suggested in \cite{jarlsprl}, $M$ is the dimensionless mass matrix $M=m/\Lambda$, where $\Lambda$ is a scale which may be chosen for convenience. A natural choice would be $\Lambda=m_t$ for the up-type or $\Lambda'=m_b$ for the down-type quarks, but since we in principle allow arbitrary values for the quark masses we leave $\Lambda$ arbitrary. Clearly, $U_L$ and $U_R$ are only defined up to simultaneous left multiplication by a diagonal unitary matrix
\ben
U_L\rightarrow AU_L\,,\quad U_R\rightarrow AU_R\,,\quad A\in U(1)^3
\een
which reduces the number of (real) parameters from 21 to 18. There are additional discrete ambiguities, corresponding to a permutation or change of sign of the elements of $D$, given by elements of the group $\frak{S}_3\times\mathbb{Z}_2^3$, where $\frak{S}_3$ is the symmetric group of three elements. (We will see shortly that the measure vanishes whenever elements of $D$ coincide up to sign, so we can restrict to matrices with $D_1^2\neq D_2^2\neq D_3^2\neq D_1^2$.) Hence we can identify the relevant subspace of the space of $3\times 3$ complex matrices with $\mathbb{R}^3\times (U(1)^3\times\frak{S}_3\times \mathbb{Z}_2^3)\backslash(U(3)\times U(3))\simeq\mathbb{R}_+^3\times (U(1)^3\times\frak{S}_3)\backslash(U(3)\times U(3))$. However, since all expressions will only involve absolute values of the elements of $D$ and $D'$ (precisely due to the $\mathbb{Z}_2^3$ symmetry), we will integrate over all of $\mathbb{R}^3$ for simplicity. This only leads to an extra factor 8 which drops out in expectation values. The metric we will use to determine a measure is
\ben
ds^2 = \Tr (dM\cdot dM^{\dagger})\,,
\een
which clearly is invariant under $M\rightarrow OMO'$ for $O,O'\in U(3)$. To evaluate this, use
\bea
dM & = & -U_L^{\dagger} dU_L U_L^{\dagger} D U_R + U_L^{\dagger} dD U_R + U_L^{\dagger} D dU_R\,,\nn
\\ dM^{\dagger} & = & -U_R^{\dagger} dU_R U_R^{\dagger} D U_L + U_R^{\dagger} dD U_L + U_R^{\dagger} D dU_L
\eea
and $[D,dD]=0$ to obtain
\ben
\Tr (dM\cdot dM^{\dagger}) = \Tr(dD\cdot dD) - \Tr(D^2(dU_L U_L^{\dagger})^2) - \Tr(D^2(dU_R U_R^{\dagger})^2) + 2 \Tr (D dU_L U_L^{\dagger} D dU_R U_R^{\dagger})\,.
\een
We introduce right-invariant one-forms
\ben
dU_L U_L^{\dagger}=\im \lambda_a \tau^a_L\,,\quad dU_R U_R^{\dagger}=\im \lambda_b \tau^b_R\,,
\een
where $\lambda_1,\ldots,\lambda_8$ are the Gell-Mann matrices, and
\ben
\lambda_9=\sqrt{\frac{2}{3}}\left(\begin{matrix} 1 & 0 & 0 \cr 0 & 1 & 0 \cr 0 & 0 & 1 \end{matrix}\right)\,,
\een
so that ${\rm i}\lambda_a$ are a basis for the Lie algebra $\frak{u}(3)$. Then
\ben
ds^2 = \Tr (dM\cdot dM^{\dagger}) = \Tr(dD\cdot dD) +\Tr(D^2\lambda_{(a}\lambda_{b)})(\tau^a_L\tau^b_L+\tau^a_R\tau^b_R) - \Tr (D \lambda_a D \lambda_b)(\tau^a_L\tau^b_R+\tau^a_R\tau^b_L)\,.
\een
The only nonvanishing traces are
\ben
\Tr(D^2\lambda_1\lambda_1)=\Tr(D^2\lambda_2\lambda_2)=\Tr(D^2\lambda_3\lambda_3)=\Tr(D\lambda_3 D\lambda_3)=D_1^2+D_2^2\,,\nn
\een
\[\Tr(D^2\lambda_{(3}\lambda_{8)})=\Tr(D\lambda_{(3}D\lambda_{8)})=\frac{1}{\sqrt{3}}(D_1^2-D_2^2)\,,\quad \Tr(D^2\lambda_{(3}\lambda_{9)})=\Tr(D\lambda_{(3}D\lambda_{9)})=\sqrt{\frac{2}{3}}(D_1^2-D_2^2)\,,\]
\[\Tr(D^2\lambda_{(8}\lambda_{9)})=\Tr(D\lambda_{(8}D\lambda_{9)})=\frac{\sqrt{2}}{3}(D_1^2+D_2^2-2D_3^2)\,,\]
\[\Tr(D^2\lambda_4\lambda_4)=\Tr(D^2\lambda_5\lambda_5)=D_1^2+D_3^2\,,\quad \Tr(D^2\lambda_6\lambda_6)=\Tr(D^2\lambda_7\lambda_7)=D_2^2+D_3^2\,,\]
\[\Tr(D^2\lambda_8\lambda_8)=\Tr(D\lambda_8 D\lambda_8)=\frac{1}{3}(D_1^2+D_2^2+4D_3^2)\,,\quad \Tr(D^2\lambda_9\lambda_9)=\Tr(D\lambda_9 D\lambda_9)=\frac{2}{3}(D_1^2+D_2^2+D_3^2)\,,\]
\[\Tr(D\lambda_1 D\lambda_1)=\Tr(D\lambda_2 D\lambda_2)=2D_1 D_2\,,\]
\ben
\Tr(D^2\lambda_4\lambda_4)=\Tr(D^2\lambda_5\lambda_5)=2 D_1 D_3\,,\quad \Tr(D^2\lambda_6\lambda_6)=\Tr(D^2\lambda_7\lambda_7)=2 D_2 D_3\,.
\een
The metric can be written in the form
\bea
ds^2 & = & dD_1^2+dD_2^2+dD_3^2 + \frac{1}{2}(D_1-D_2)^2(\tau_L^1+\tau_R^1)^2 + \frac{1}{2}(D_1+D_2)^2(\tau_L^1-\tau_R^1)^2\nonumber
\\ & & + \frac{1}{2}\left[(D_1-D_2)^2(\tau_L^2+\tau_R^2)^2 + (D_1+D_2)^2(\tau_L^2-\tau_R^2)^2\right.\nn
\\ & & \left.+ (D_1-D_3)^2(\tau_L^4+\tau_R^4)^2 + (D_1+D_3)^2(\tau_L^4-\tau_R^4)^2\right]\nonumber
\\ & & + \frac{1}{2}\left[(D_1-D_3)^2(\tau_L^5+\tau_R^5)^2 + (D_1+D_3)^2(\tau_L^5-\tau_R^5)^2\right.\nn
\\ & & \left.+ (D_2-D_3)^2(\tau_L^6+\tau_R^6)^2 + (D_2+D_3)^2(\tau_L^6-\tau_R^6)^2\right]\nonumber
\\ & & + \frac{1}{2}(D_2-D_3)^2(\tau_L^7+\tau_R^7)^2 + \frac{1}{2}(D_2+D_3)^2(\tau_L^7-\tau_R^7)^2\nonumber
\\ & & + 2 D_1^2 \left(\frac{1}{\sqrt{2}}(\tau_L^3 - \tau_R^3) + \frac{1}{\sqrt{6}}(\tau_L^8 - \tau_R^8)+\frac{1}{\sqrt{3}}(\tau_L^9 - \tau_R^9)\right)^2 \nonumber
\\ & & + 2 D_2^2 \left(-\frac{1}{\sqrt{2}}(\tau_L^3 - \tau_R^3) + \frac{1}{\sqrt{6}}(\tau_L^8 - \tau_R^8)+\frac{1}{\sqrt{3}}(\tau_L^9 - \tau_R^9)\right)^2\nn
\\ & & + 2 D_3^2 \left(\sqrt{\frac{2}{3}}(\tau_L^8 - \tau_R^8)+\frac{1}{\sqrt{3}}(\tau_L^9 - \tau_R^9)\right)^2.
\eea
The fact that the metric only depends on $\tau_L^3-\tau_R^3$ etc., and not on $\tau_L^3+\tau_R^3$ etc., again reflects the $U(1)^3$ that has to be factored out. As is easy to show, the volume form is proportional to 
\[(D_1^2-D_2^2)^2(D_1^2-D_3^2)^2(D_2^2-D_3^2)^2 |D_1 D_2 D_3|\,dD_1\wedge dD_2\wedge dD_3\wedge \tau_L^1\wedge \tau_R^1\wedge\ldots\wedge(\tau_L^8-\tau_R^8)\wedge(\tau_L^9-\tau_R^9)\,.\]
Note that this expression only depends on the absolute values of $D_i$, as expected. Since the range of the $D_i$ is infinite, integration over these coordinates will give an infinity, so that a function decaying sufficiently fast for large $|D_i|$ is introduced. A natural and simple choice is a Gaussian.

Since there are actually two integrations over the space of mass matrices, corresponding to two mass matrices for the up-type and down-type quarks, the general expression for the measure considered in \cite{paper,prl}, where $M$ was assumed to be Hermitian,  was
\[DM\,DM'\,\exp(-\Tr(M^2 A))\exp(-\Tr((M')^2 A))\,,\]
where $DM$ was the natural measure on the space of $3\times 3$ Hermitian matrices, with $A$ and $A'$ Hermitian with non-negative eigenvalues and commuting. $A$ and $A'$ could then be diagonalized by the transformation $U\rightarrow UW$, $U'\rightarrow U'W$ without changing any measurable quantities. The Gaussian broke the symmetry of the measure from invariance under $(U(3)\times U(3))$ to invariance under $(U(1)^3\times U(1)^3)$. 

The analogous procedure in the case of general complex mass matrices is to use the measure
\[DM\,DM'\,\exp(-\Tr(MM^{\dagger}A+M'(M')^{\dagger}A'))=DM\,DM'\,\exp(-\Tr(U_L^{\dagger}D^2 U_L A+(U'_L)^{\dagger}(D')^2 U'_L A'))\,,\]
so that the measure is still invariant under the right $U(3)$ action on complex matrices, but the invariance under the left $U(3)$ action is again broken to $U(1)^3$. This symmetry breaking is necessary to obtain a distribution that can reproduce the observed quark masses. As before, we assume $A$ and $A'$ to be Hermitian with non-negative eigenvalues and commuting and use the redefinitions $U_L\rightarrow U_LW_L$, $U_L'\rightarrow U_L'W_L$ to diagonalise $A$ and $A'$. 

Since the important measurable quantity involving the unitary matrices is the CKM matrix $V=U_L U_L'^{\dagger}$, everything is independent of the $U_R$ parameters and we integrate over these, obtaining a constant which is irrelevant in the averaging process. 

We are left with integrating over the space of possible $U_L$, the coset $U(1)^3 \backslash U(3) = U(1)^2 \backslash SU(3)$ (both the Gaussian and the CKM matrix are invariant under left multiplication of $U_L$ by an element of $U(1)^3$), and the volume form is proportional to
\[(D_1^2-D_2^2)^2(D_1^2-D_3^2)^2(D_2^2-D_3^2)^2 |D_1 D_2 D_3|\,dD_1\wedge dD_2\wedge dD_3\wedge \tau_L^1\wedge \tau_L^2\wedge \tau_L^4 \wedge \tau_L^5\wedge\tau_L^6\wedge\tau_L^7\,.\]
In terms of the coordinates on $U(1)^2\backslash SU(3)$ introduced in section 2, the wedge product of right-invariant forms gives the usual bi-invariant measure on $SU(3)$, so that we finally get
\ben
DM = \left(\prod_{i<j}(D_i^2-D_j^2)^2\right) |D_1 D_2 D_3|\,\sin 2x \cos^3 y \sin y \sin 2z\,dD_1\, dD_2\, dD_3\,dx\,dy\,dz\,dw\,dr\,dt.
\label{su3meas}
\een

We also have to take the discrete $\frak{S}_3$ symmetry into account, integrating only over one sixth of the homogeneous space $U(1)^2\backslash SU(3)$, corresponding to
\ben
0\le y\le \arctan(\sin x)\,,\quad 0\le x\le\frac{\pi}{4}\,.
\een
This restriction amounts to removing unitary matrices that permute the elements of $D$ and hence to fixing an ordering.

Comparing to the case of Hermitian matrices \cite{prl}, the measure involves higher powers of elements of $D$. Therefore, we would expect a stronger influence of the quark mass hierarchy on expectation values of (powers of) $J$, namely a stronger suppression of large values of $J$. This naive expectation will be confirmed in the next section.

\section{Results for $J$}

The calculations go through exactly as in the case of the measure used in \cite{paper}, and our analysis will be completely analogous. We will obtain analytical approximations of the relevant integrations whose validity is then verified by numerical integration (using {\sc Mathematica}).

We want to compute the expectation value for $J^2$ in the given probability distribution (all odd powers of $J$ average to zero),
\ben
\langle J^2\rangle = \frac{\int DM\,DM'\,e^{-\Tr(MM^{\dagger}A)-\Tr(M'(M')^{\dagger}A')} J^2(M,M') }{\int DM\,DM'\,e^{-\Tr(MM^{\dagger}A)-\Tr(M'(M')^{\dagger}A')}}\,,
\label{jint}
\een
where $DM$ is given in (\ref{su3meas}) and $DM'$ is the same expression in terms of primed variables. It should be clear that multiplying $A$ or $A'$ by a (non-negative) constant is the same as rescaling the arbitrary scales $\Lambda$ and $\Lambda'$ used in defining $D_i$ or $D_i'$. Hence we may assume, without loss of generality, the form
\ben
A = \left(\begin{matrix} 1 & 0 & 0 \\ 0 & \mu_c^{-2} & 0 \\ 0 & 0 & \mu_u^{-2}\end{matrix}\right)\,,\quad A' = \left(\begin{matrix} 1 & 0 & 0 \\ 0 & \mu_s^{-2} & 0 \\ 0 & 0 & \mu_d^{-2}\end{matrix}\right)\,.
\label{aform}
\een 
We have introduced four dimensionless parameters $\mu_c,\mu_u,\mu_s$, and $\mu_d$ that we are free to choose so as to reproduce the observed values of the quark masses as expectation values in our distribution.\footnote{We emphasise that we do not make any predictions about quark masses, but use them as an input to modify the probability distribution on the space of all mass matrices.} On dimensional grounds alone, we expect that we will have to set $\mu_c\approx m_c/m_t$ etc. In particular, we will have $\mu_u\ll 1$ and $\mu_d\ll 1$. The precise relation between the $\mu$ parameters and expectation values for (squared) quark masses will be determined shortly.

Now $J^2$, written in terms of coordinates on $M$ and $M'$, is a very complicated expression. However, since
\ben
\Tr(MM^{\dagger}A) = \sum_{ij}D_i^2 A_{j} |U_L|_{ij}^2 = \frac{D_1^2}{\mu_u^2}\sin^2 y + \ldots \,,\quad \Tr(M'(M')^{\dagger}A') = \frac{(D_1')^2}{\mu_d^2}\sin^2 y' + \ldots \,,
\een
where we assume $\mu_u\ll 1$ and $\mu_d\ll 1$ and $A_j$ denote the diagonal elements of $A$, the integrand in the numerator and denominator of (\ref{jint}) is negligibly small unless $y$ and $y'$ are close to zero. We therefore use the approximation $\sin y\approx y$ in the measure $DM$, setting $y=y'=0$ in the remaining part of the integrand. The integration over $y$ and $y'$ can be easily performed, and everything is independent of $w$ and $w'$ which may hence be dropped as integration variables. One is left with the integral
\ben
\langle J^2 \rangle \approx \frac{\int dD \,dD' \int d^4 x\, d^4 x'\sin 2x \sin 2z \sin 2x' \sin 2z' \left(e^{-\Tr(D^2 UAU^{\dagger}+(D')^2 U'A'{U'}^{\dagger})}\,J^2(U,U')\right)\big|_{y=y'=0}}{\int dD \, dD' \int d^4 x\, d^4 x'\sin 2x \sin 2z \sin 2x' \sin 2z' \left(e^{-\Tr(D^2 UAU^{\dagger}+(D')^2 U'A'{U'}^{\dagger})}\right)\big|_{y=y'=0}}\,,
\label{newjint}
\een
where
\ben
\int d^4 x\equiv \int\limits_0^{\pi/4}dx\int\limits_0^{\pi/2}dz \int\limits_0^{2\pi}dr\int\limits_0^{2\pi}dt
\een
and similarly for $\int d^4 x'$. The shorthand $dD$ denotes the measure over the $D_i$, namely
\ben
dD = \left(\prod_{i<j}(D_i^2-D_j^2)^2\right) |D_1 D_2 D_3|\,dD_1\, dD_2\, dD_3\,.
\een
It is possible to analytically integrate over both copies of $\R^3$ in (\ref{newjint}), using
\bea
f_{\xi_1\xi_2\xi_3} & := &\int\limits_{-\infty}^{\infty}dD_1\int\limits_{-\infty}^{\infty}dD_2\int\limits_{-\infty}^{\infty}dD_3\,(D_1^2-D_2^2)^2(D_1^2-D_3^2)^2(D_2^2-D_3^2)^2 |D_1 D_2 D_3| e^{-\xi_1 D_1^2-\xi_2D_2^2-\xi_3D_3^2}\nn
\\& = &\frac{24}{\xi_1^{5}\xi_2^{5}\xi_3^{5}}\left(2(\xi_1^2\xi_2^2+\xi_1^2\xi_3^2+\xi_2^2\xi_3^2)(\xi_1^2+\xi_2^2+\xi_3^2-\xi_1\xi_2-\xi_1\xi_3-\xi_2\xi_3)\right.\nn
\\& & \left.-3\xi_1\xi_2\xi_3(\xi_1^3+\xi_2^3+\xi_3^3-\xi_2^2\xi_3-\xi_3^2\xi_1-\xi_1^2\xi_2-\xi_3^2\xi_2-\xi_1^2\xi_3-\xi_2^2\xi_1)-8\xi_1^2\xi_2^2\xi_3^2\right)\,.
\eea

The explicit expression for $J$ at $y=y'=0$ is
\bea
J(U,U')\big|_{y=y'=0} & = &\frac{1}{4}s_{2x} s_{2x'} \left\{c^2_{z'} s^3_{z} s_{z'}\sin(3 \hat{r} + \hat{t})  + c^3_{z} c_{z'} s^2_{z'}\sin(3 \hat{r} - \hat{t}) \right.\nn
\\& &  - c^2_{z} s_{z} s_{z'} (c^2_{z'} \left[\sin(3 \hat{r} + \hat{t}) + \sin(3\hat{r} -3\hat{t})\right] - s^2_{z'}\sin(3\hat{r} + \hat{t}) )\nn
\\& & \left. + c_{z} c_{z'} s^2_{z} (c^2_{z'} \sin(3\hat{r} - \hat{t}) -  s^2_{z'}\left[\sin(3 \hat{r} + 3\hat{t}) + \sin(3 \hat{r} - \hat{t})\right])\right\}
\label{jexp}
\eea
where $s_x=\sin x, c_{z'}=\cos z'$, etc., $\hat r= r-r'$, and $\hat t=t-t'$. Integrating (\ref{jexp}) over $r,r',t$, and $t'$ indeed gives zero, which is why we choose to use $J^2$.

To fix the parameters appearing in the matrices $A$ and $A'$, we first observe that expectation values for squared mass matrices, again in the approximation $y\ll 1$, take the relatively simple form
\ben
\langle D_1^2 \rangle \approx \frac{\int_{\R^3} dD\, D_1^2 \int\limits_0^{\pi/4} dx \int\limits_0^{\pi/2} dz\,\sin 2x \sin 2z  \left(e^{-\Tr(D^2 UAU^{\dagger})}\right)\big|_{y=0}}{\int_{\R^3} dD\,  \int\limits_0^{\pi/4} dx \int\limits_0^{\pi/2} dz\,\sin 2x \sin 2z  \left(e^{-\Tr(D^2 UAU^{\dagger})}\right)\big|_{y=0}}\,.
\een
The denominator is explicitly
\ben
I_D:=\int\limits_0^{\pi/4} dx \int\limits_0^{\pi/2} dz\,\sin 2x \sin 2z \,f_{\xi_1\xi_2\xi_3}\,,
\label{denom}
\een
where
\bea
& \xi_1=A_1 \cos^2 z + A_2 \sin^2 z\,,\quad \xi_2=A_1 \cos^2 x \sin^2 z + A_2 \cos^2 x \cos^2 z + A_3 \sin^2 x\,,\nn
\\& \xi_3=A_1 \sin^2 x \sin^2 z + A_2 \sin^2 x \cos^2 z + A_3 \cos^2 x\,,
\eea
with $A_3 \gg A_2 \gg A_1$. The integral is dominated by very small $x$ and $z$ (we cannot have $x=\frac{\pi}{2}$), and we can approximate $I_D$ well by only keeping the terms of leading order in $x$ and $z$ in the trigonometric functions, and 
\ben
f_{\xi_1\xi_2\xi_3} \approx \frac{24}{\xi_1^5\xi_2^5\xi_3^5}\times 2\xi_3^4\xi_2^2\approx \frac{48}{(A_1+A_2 z^2)^5 (A_2 + A_3 x^2)^3 A_3}\,,
\een
which are the leading terms (as we shall see, the combination $A_3 x^2$ is effectively of order $A_2$ etc.):
\bea
I_D & \approx & \frac{48}{A_3} \int\limits_0^{\pi/4} dx \int\limits_0^{\pi/2} dz\,4 x z \,(A_1+A_2 z^2)^{-5} (A_2 + A_3 x^2)^{-3}\nn
\\ & \approx &\frac{48}{A_3} \int\limits_0^{\infty} dX \int\limits_0^{\infty} dZ\,(A_1+A_2 Z)^{-5} (A_2 + A_3 X)^{-3}\nn
\\ & = &\frac{48}{A_3}\cdot\frac{1}{4A_2 A_1^4}\cdot\frac{1}{2A_3 A_2^2} = \frac{6}{A_1^4 A_2^3 A_3^2}\,.
\eea
This is very well reproduced by numerical calculations. Similarly, we find
\bea
I_D\langle D_1^2 \rangle & \approx & \frac{240}{A_3} \int\limits_0^{\infty} dX \int\limits_0^{\infty} dZ\,(A_1+A_2 Z)^{-6} (A_2 + A_3 X)^{-3}\nn
\\ & = & \frac{240}{A_3}\cdot\frac{1}{5A_2 A_1^5}\cdot\frac{1}{2A_3 A_2^2} = \frac{24}{A_1^5 A_2^3 A_3^2}\,,
\eea
hence
\ben
\langle D_1^2 \rangle \approx \frac{4}{A_1}\,.
\een
Redoing the same calculation for $D_2$ and $D_3$ gives
\ben
\langle D_2^2 \rangle \approx \frac{2}{A_2}\,,\quad \langle D_3^2 \rangle \approx \frac{1}{A_3}\,.
\een
Using the form (\ref{aform}) for $A$, this is
\ben
\langle D_1^2\rangle\approx 4\,,\quad \langle D_2^2 \rangle \approx 2\mu_c^2\,,\quad \langle D_3^2 \rangle \approx \mu_u^2\,.
\een
If $\langle D_1^2\rangle$ is to reproduce the squared top quark mass in units of $\Lambda$, our reference scale for the up-type quarks must be $\Lambda=\frac{1}{2}m_t$. Then setting $m_c^2/\Lambda^2=2\mu_c^2$ determines $\mu_c$, etc. Apart from numerical prefactors of order one, the $\mu$ parameters indeed correspond to the quark masses one wants to reproduce in the probability distribution. Note that factoring out the $\frak{S}_3$ above corresponds to fixing an ordering of the quark masses, so that it is possible to obtain unequal expectation values for $D_1,D_2$ and $D_3$. Integrating over the whole of $U(1)^2\backslash SU(3)$ would mean to also include permutations, so that necessarily $\langle D_1^2\rangle=\langle D_2^2\rangle=\langle D_3^2\rangle$.

Again, because we have to consider the dependence of masses on the energy scale in quantum field theory, described by the renormalization group, there is some ambiguity in what is meant by the ``quark masses" we want to reproduce. Following \cite{rosner}, for example, we take all the quark masses evolved to the scale of the $Z$ boson mass. These are given in \cite{massref}:
\bea
&&(m_u,m_c,m_t)=(1.27_{-0.42}^{+0.50}\;{\rm MeV},\; 0.619 \pm 0.084\;{\rm GeV},\;171.7 \pm 3.0\;{\rm GeV})\,; \nn
\\&& (m_d,m_s,m_b)=(2.90_{-1.19}^{+1.24}\;{\rm MeV},\; 55_{-15}^{+16}\;{\rm MeV},\;2.89 \pm 0.09\;{\rm GeV})\,.
\label{qmass}
\eea
We use the central values
\bea
&&(m_u,m_c,m_t):=(1.27\;{\rm MeV},\; 0.619\;{\rm GeV},\;171.7\;{\rm GeV})\,; \nn
\\&& (m_d,m_s,m_b):=(2.9\;{\rm MeV},\; 55\;{\rm MeV},\;2.89\;{\rm GeV})\,.
\label{cvalues}
\eea

The mass scales $\Lambda$ and $\Lambda'$ are now fixed by setting $\langle D_1^2 \rangle=(m_t/\Lambda)^2$ and $\langle (D'_1)^2 \rangle=(m_b/\Lambda')^2$. By comparing the results obtained by numerical integration with the values we want to reproduce, we can then fix the parameters $\mu_c,\mu_u,\mu_s$ and $\mu_d$.

In the case of the positively charged top, charm and up quarks, which exhibit a more extreme quark mass hierarchy, we find that numerical calculations (using {\sc Mathematica}) reproduce the results we have obtained analytically very well. For the negatively charged quarks, we find numerically that we have to use different relative factors to reproduce the observed masses. Comparing the numerical results with (\ref{cvalues}), we fix the parameters appearing in $A$ and $A'$ to
\bea
\mu_c^2 = 2\left(\frac{m_c}{m_t}\right)^2\approx 2.60\times 10^{-5}\,,\quad \mu_u^2 = 4\left(\frac{m_u}{m_t}\right)^2\approx 2.19\times 10^{-10}\,,\nn
\\ \mu_s^2 = \left(\frac{m_s}{m_b}\right)^2\approx 3.62\times 10^{-4}\,,\quad \mu_d^2 = 4\left(\frac{m_d}{m_b}\right)^2\approx 4.03\times 10^{-6}\,.
\label{muvalues}
\eea

In order to obtain an analytical expression for expectation value of $J^2$, the next approximation is that the main contribution to the integral (\ref{newjint}) will come from small $z$. This again is seen by writing out $\Tr(MM^{\dagger} A)$ and using the mass hierarchy. We only take the term in (\ref{jexp}) that is non-zero at $z=0$ into account, setting $\sin 2z\approx 2z$ in the measure.

Averaging over $r,t,r'$,and $t'$ gives a factor of 1/2, as one might have expected, and therefore we use
\ben
J^2_{{\rm small}\;z}:=\frac{1}{2}\sin^2 x\cos^2 x\sin^2 x'\cos^2 x'\cos^2 z'\sin^4 z'
\een
for our calculations. Within this approximation for $J$, still taking $f_{\xi_1\xi_2\xi_3} \approx 48\xi_1^{-5}\xi_2^{-3}\xi_3^{-1}$, the numerator of (\ref{newjint}) is the product (using again that only small $z$ contributes)
\bea
I_N & \approx &1152\times \int\limits_0^{\pi/2} dz \, \frac{2z}{(A_1 + A_2 z^2)^{5/2}} \times \int\limits_0^{\pi/2} dz' \, \frac{\sin 2z' \cos^2 z'\sin^4 z'}{(A_1 \cos^2 z'+ A_2 \sin^2 z')^5} \nn
\\ & & \times \int\limits_0^{\pi/4} dx \,\frac{\sin 2x\,\sin^2 x\,\cos^2 x}{(A_2 \cos^2 x + A_3 \sin^2 x)^3 (A_3 \cos^2 x + A_2 \sin^2 x)}\nn
\\ & & \times \int\limits_0^{\pi/4} dx' \,\frac{\sin 2x'\,\sin^2 x'\,\cos^2 x'}{(A'_2 \cos^2 x' + A'_3 \sin^2 x')^3 (A'_3 \cos^2 x' + A'_2 \sin^2 x')}
\eea
The first two factors are $1/(4A_1^4 A_2)$ and $1/(12(A_1')^2(A_2')^3)$, respectively; for the other two (which have the same form) we change variables to $X=\cos^2 x$ to obtain
\ben
\int\limits_{1/2}^{1} dX \,\frac{X(1-X)}{(A_2 X + A_3 (1-X))^3 (A_3 X + A_2 (1-X))}= \frac{1}{2A_3^3 A_2}\left(1-\frac{2A_2}{A_3}+O\left(\left(A_2/A_3\right)^2\right)\right)\,;
\een
the expressions for the denominator are similar but simpler. Putting everything together, we obtain the approximation to lowest order in quark mass ratios
\ben
\langle J^2_{{\rm small}\;z}\rangle \approx \frac{1}{6}\frac{A_2 (A_1')^2}{A_3 A_2' A_3'} = \frac{4}{3}\frac{m_s^2 m_u^2 m_d^2}{m_b^4 m_c^2}\,,
\label{japprox}
\een
where the numerical factors appearing in the last line come from the different factors chosen in (\ref{muvalues}). Note that the top quark mass does not appear in this approximate result. This compares with the scaling behaviour obtained in \cite{paper} (for Hermitian mass matrices),
\ben
\langle J^2_{{\rm small}\;z}\rangle \sim \frac{m_s^2 m_u m_d}{m_b^3 m_c}.
\een

For numerical calculations we use both the simplified expression $J^2_{{\rm small}\;z}$ and the expression for $J$ given in (\ref{jexp}). We find that for the first quantity, the numerically evaluated expectation value, $\langle J^2_{{\rm small}\;z}\rangle\approx 1.89\times 10^{-15}$, is about $94\%$ of (\ref{japprox}), and the numerical result for $\langle J^2 \rangle$ (taken at $y=y'=0$) is
\ben
\langle J^2 \rangle \approx 2.07\times 10^{-15}\,,
\een
which gives
\ben
\Delta J = \sqrt{\langle J^2 \rangle} \approx 4.55 \times 10^{-8}
\een
which is now almost three orders of magnitude {\it smaller} than the observed value (\ref{jobserved}). Assuming a Gaussian distribution peaked at zero, we get
\ben
P(|J|\le 10^{-7})\approx 97\%\,,
\een 
When the measure presented here is used, there seems to be extreme fine-tuning in $J$, but now we would say that one observes unnaturally {\it large} CP violation! This result may look surprising, given that the maximal value for $J$ is around 0.1 and the observed value just $3\times 10^{-5}$, but it shows how strongly the quark mass hierarchy suppresses large values of $J$ in our distribution.

\section{Summary and Discussion}

We have tried to estimate the naturalness of the observed magnitude of $CP$ violation in the electroweak theory under the assumption that there is a left-right symmetry which implies that the quark mass matrices can not in general be taken to be Hermitian. We have constructed a probability distribution on the space of $3\times 3$ complex matrices which takes into account the geometrical structure of this space, but also includes a Gaussian factor which makes the total volume finite and leads to expectation values for quark masses that can reproduce the observed values if four free parameters are fitted accordingly. While this is a choice we make, and all results depend on this choice, our measure is the combination of a maximally symmetric measure, invariant under a redefinition of a complex matrix by left or right multiplication by unitary matrices, and a Gaussian incorporating the observed values of the quark masses. We would have to make additional rather strong assumptions to motivate a different choice of measure that would differ appreciably from this simple construction. It may well be that such assumptions are justified by the underlying mechanism determining the mass matrices, but we do not know of such a mechanism yet. Furthermore, such a simple choice for the measure was shown in \cite{paper}, where only Hermitian mass matrices were considered, to lead to expectation values for the Jarlskog invariant $J$ that make the observed value appear typical. 

The conclusion for general complex mass matrices, as shown here, is rather different. Using the given probability distribution, one would now expect $J$ to be about three orders of magnitude smaller than the observed value. Hence, there is a fine-tuning problem: Without further assumptions, a fundamental theory leading to a left-right symmetric electroweak sector at low energy should generically be expected to reproduce very weak $CP$ violation. Invoking the principle of Occam's razor, ``{\sc entia non sunt multiplicanda praeter necessitatem}," we would like to conclude that, only looking at possible explanations for the magnitude of $CP$ violation in the electroweak sector, the standard model should be preferred to left-right symmetric extensions such as Pati-Salam: In the latter one needs additional assumptions on the fundamental parameters that resolve the issue of observing ``unnaturally large" $CP$ violation, that are not necessary in the standard model, or any extension of it that allows a restriction to Hermitian mass matrices only.

Although we have tried to argue that our results are independent of renormalization group flow since the relevant quantities do not run strongly with energy scale, there is another subtle issue: The low-energy limit of a left-right symmetric extension with non-Hermitian mass matrices would still be the standard model, where mass matrices can be assumed to be Hermitian, leading us back to the measure considered in \cite{paper, prl}. It would be desirable to incorporate this dependency of the assumptions one has used to construct the measure on energy scale into the analysis, namely to use a measure which depends on energy scale also. A starting point would be a quantification of ``non-Hermiticity" that could then flow from zero at low energies to some non-zero value at high energies. At present these ideas are however somewhat vague, so that we will have to leave them to exploration in future work.

\section*{Acknowledgments}
The groundwork for the simple calculations presented here was laid in the previous papers \cite{paper,prl}, and I thank my collaborators in this previous work for many fruitful discussions and suggestions. I am supported by EPSRC and Trinity College, Cambridge. I should also thank the referee for suggestions that hopefully led to an improvement of presentation.

\end{document}